\documentclass{emulateapj}

\usepackage{graphicx}
\usepackage{epsfig}
\usepackage{times}
\usepackage[]{natbib}
\usepackage{url}
\usepackage{color}
\usepackage{amssymb,amsmath}

\newcommand{\jcph}{{Journal of Comp. Phys.}}

\shorttitle{CME-like eruptions driven by dynamic flux emergence.}
\shortauthors{Archontis et al.}

\begin{document}

\title{Recurrent explosive eruptions and the ``sigmoid-to-arcade'' transformation in the Sun driven by dynamical magnetic flux emergence.}

%\author{V. Archontis\altaffilmark{1,2}}
%\author{A.W. Hood\altaffilmark{2}}
%\and
%\author{K. Tsinganos\altaffilmark{3}}

\author{V. Archontis\altaffilmark{1,2},
A.W. Hood\altaffilmark{2} and K.Tsinganos\altaffilmark{3}}

\altaffiltext{1}{Section of Astrophysics, Astronomy and Mechanics, Department of Physics, University of Athens, Panepistimiopolis 15784, Athens, Greece}
\altaffiltext{2}{School of Mathematics and Statistics, St. Andrews University, St. Andrews, KY169SS, UK}
\altaffiltext{3}{National Observatory of Athens, Lofos Nymphon, Thissio, Athens, Greece}

\email{va11@st-andrews.ac.uk}

%% Mark off your abstract in the ``abstract'' environment. In the manuscript
%% style, abstract will output a Received/Accepted line after the
%% title and affiliation information. No date will appear since the author
%% does not have this information. The dates will be filled in by the
%% editorial office after submission.

\begin{abstract}
We report on three-dimensional (3D) MHD simulations of recurrent mini CME-like eruptions in a small active region (AR), 
which is formed by the {\it dynamical} emergence of a twisted (not kink unstable) flux tube from the solar interior. 
The eruptions are developed as a result of repeated formation and expulsion of new flux ropes due to continous emergence 
and reconnection of sheared fieldlines along the polarity inversion line (PIL) of the AR. The acceleration of the eruptions 
is triggered by tether-cutting reconnection at the current sheet underneath the erupting field. We find that each explosive eruption is followed 
by reformation of a sigmoidal structure and a subsequent ``sigmoid-to-flare arcade'' transformation in the AR. These results might have 
implications for recurrent CMEs and eruptive sigmoids/flares observations and theoretical studies.
\end{abstract}

%% Keywords should appear after the \end{abstract} command. The uncommented
%% example has been keyed in ApJ style. See the instructions to authors
%% for the journal to which you are submitting your paper to determine
%% what keyword punctuation is appropriate.

\keywords{Sun: activity --- Sun: magnetic topology}

\section{Introduction}
In the active Sun, the emergence of magnetic flux from the solar interior can lead to the formation of ARs 
\citep[e.g.][]{zwaan85}, which are often associated with coronal mass ejections (CMEs), flaring activity and 
other dynamic events \citep[see e.g.][]{vandriel95,schrijver09,arcr12}.
In adittion, observational studies \citep[see e.g.][]{hudson98,canfield99} have revealed that
ARs with S-shaped morphology are the most favourables candidates for eruptive activity in the Sun.
The term ``sigmoid'' has been used by \cite{rust96} to denote the overal (forward or reverse) 
S-like shape structure of an AR.
Observations \citep[see e.g.][]{sterling00,pevtsov02,liu10} have also shown that eruptive ARs, which initially 
display a degree of twist adopting a sigmoidal 
shape, evolve into a post-eruption flare arcade that consists of fieldlines with a cusp-like shape. This process is 
known as ``sigmoid-to-arcade'' evolution and it could happen repeatedly \citep[e.g.][]{gibson02,nitta01} in some ARs: a bright (EUV/X-ray) sigmoid is reformed 
after each CME(-like) eruption, which is accompanied by a hot arcade underneath it. 

Numerical experiments have demonstrated the importance of magnetic flux emergence on driving CME-like eruptions associated with 
flaring activity in emerging flux regions (EFRs) \citep[e.g.][]{shibata_r11, kusano12}. Simulations have also shown the formation of sigmoidal structures 
in EFRs in conjuction 
with the (partial) eruption of magnetic flux ropes \citep[e.g.][]{gibson06,arc09}. The onset of recurrent eruptions has been studied in the context 
of a break-out magnetic scenario \citep[e.g.][]{devore08} and in a smimilar manner in EFRs 
(e.g. reconnection between the emerging and a pre-existing magnetic field, \citep[e.g.][]{arc08,mactaggart09}). 
The repeated ``sigmoid-to-arcade'' evolution associated with CMEs has been shown by the kinematically driven (quasi-static) emergence of a highly 
twisted magnetic torus into a preexisting potential coronal field \citep[]{chat13}. In this Letter, we present a 3D MHD experiment 
of the onset of recurrent CME-like eruptions driven by the {\it dynamical} emergence of a horizontal twisted flux tube in a highly 
stratified atmosphere. We find that the ``sigmoid-to-arcade'' process occurs naturally due to reconnection of sheared fieldlines during the eruptions.

\section{The model}
%__________________________________________________________________________________________________
We solve the MHD equations in 
Cartesian geometry, using the Lare3d code \citep[]{arber01}. Viscous and 
Ohmic heating are considered through shock viscosity and Joule dissipation. 
Uniform explicit resistivity is included, with $\eta=10^{-3}$.
Initially, the plasma is embedded into a plane-parallel hydrostatic atmosphere.
A sub-photospheric adiabatically stratified layer resides in the range ($-7.2\,\mathrm{Mm} \leq z < 0\,\mathrm{Mm}$). 
The layer above, at $0\,\mathrm{Mm}\leq z < 2.3\,\mathrm{Mm}$, which
is isothermal ($5100\,\mathrm{K}$) at the beginning and then the temperature increases with height up to $\approx 4\times10^{4}\,\mathrm{K}$, 
is mimicking the photosphere/chromosphere.
The layer at $2.3\,\mathrm{Mm}\leq z \leq 3.1\,\mathrm{Mm}$ represents the transition region.
An isothermal coronal layer ($\approx 1\,\mathrm{MK}$) is included at $3.1\,\mathrm{Mm} < z \leq 57.6\,\mathrm{Mm}$. 
The initial magnetic field is a horizontal twisted flux tube at $z_{0}=-2.1\,\mathrm{Mm}$ and 
oriented along the positive $Y$-axis. The axial field of the cylindrical tube is defined by 
\begin{equation}
B_y=B_0\,\mathrm{exp}(-r^2/R^2), \,\,\,\,\, B_\phi=\alpha\,r\,B_y, 
\end{equation}
where $R=450\,\mathrm{km}$ is the radius of the tube, $r$ is the radial distance
from the tube axis ($r^{2} = x^{2} + (z + z_{0})^{2}$) and $\alpha=0.0023\mathrm{km^{-1}}$ is the uniform twist 
around the axis of the tube, which is stable to the kink instabilty. A density deficit is applied along the axis of the tube, making 
its central part more buoyant than its footpoints:
\begin{equation}
\Delta\,\rho=[p_t(r)/p(z)]\,\rho(z)\,\mathrm{exp}\,(-y^2/\lambda^2),
\end{equation}
where $p_t$ is the pressure within the flux tube and $\lambda$ defines the length of the buoyant part of the tube. 
We use $\lambda=0.9\,\mathrm{Mm}$ and an initial field strength for the tube that corresponds to plasma ${\beta}\approx14$.
The numerical domain
is $[-32.4,32.4] \times [-32.4,32.4] \times[-7.2,57.6]\,\mathrm{Mm}$ in the longitudinal ($x$), transverse
($y$) and vertical ($z$) directions, respectively. The grid has 420 nodes in all 
directions with periodic boundary conditions in $y$. Open boundary conditions have been implemented along $x$ 
and at the top of the numerical domain, allowing outflow of plasma. 
The bottom boundary is a non-penetrating, perfectly conducting wall.

\section{Results and discussion}
%========================================================
% Figure 1
\begin{figure}[t]
\centering
\includegraphics[width=1.\linewidth]{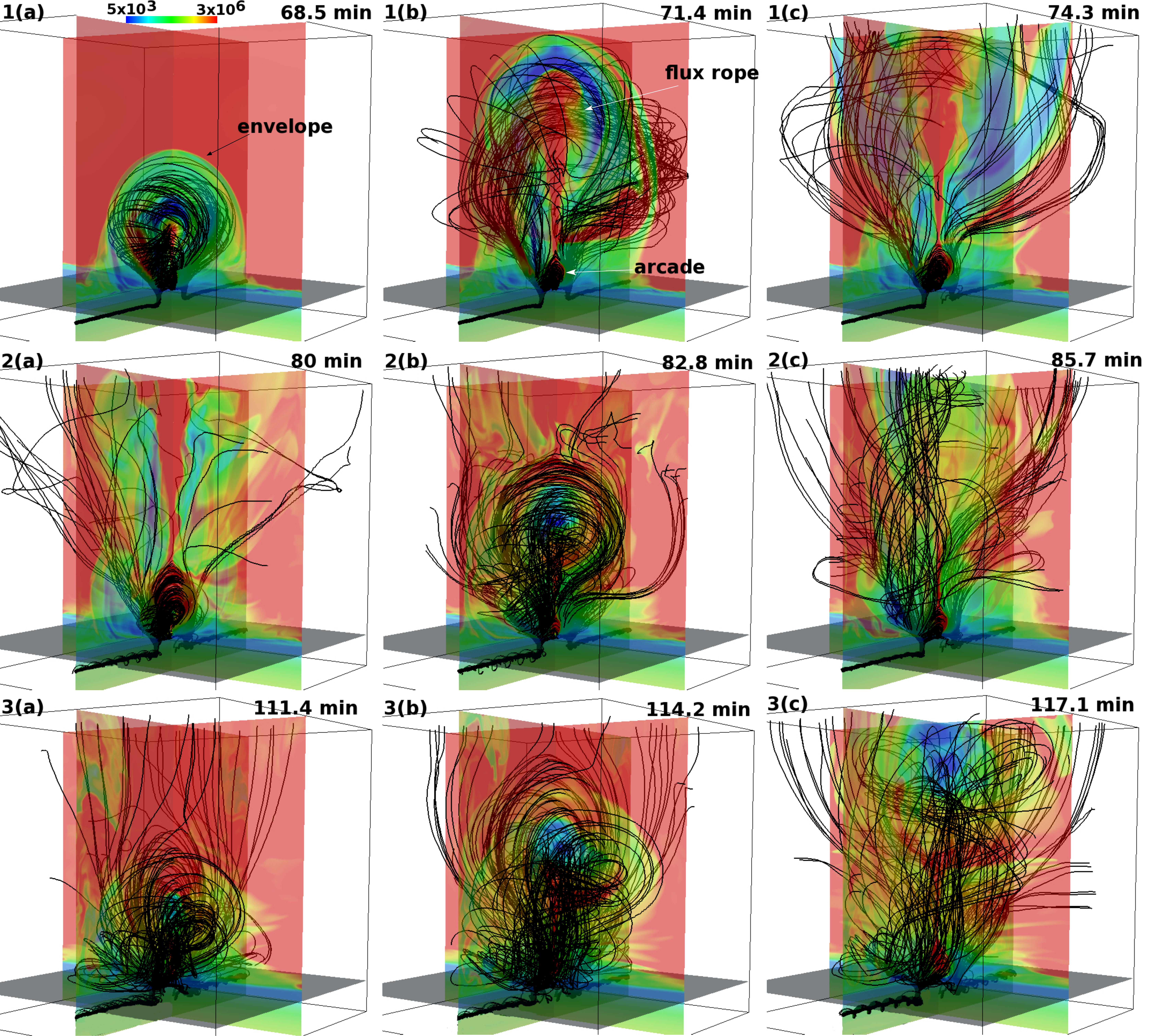}\caption{Temperature and magnetic field topology during the eruptions.
Shown are the two vertical midplanes and the horizontal slice at the base of the photosphere.}
\end{figure}
%========================================================

%========================================================
% Figure 2
\begin{figure}[t]
\centering
\includegraphics[width=0.9\linewidth]{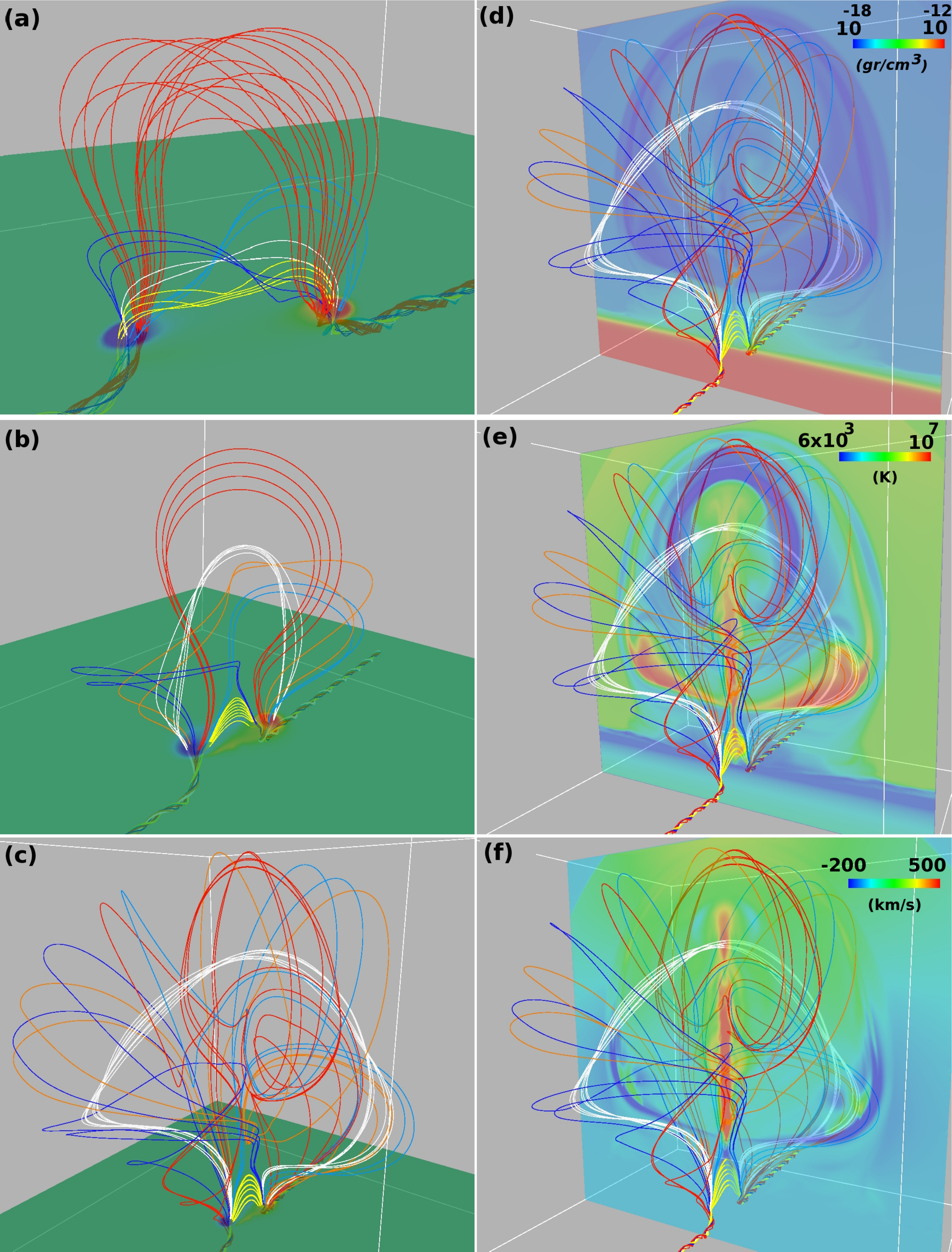}\caption{ Field line topology during the first eruption (panels (a)-(c)). 
The horizontal slice shows the Bz distribution at the photosphere (red;positive, blue;negative within the range $[-500,500]\,\mathrm{G}$.
Times are $t=62.8\,\mathrm{min}$, 
$t=68.5\,\mathrm{min}$ and $t=71.4\,\mathrm{min}$ for panels (a)-(c) respectively. Logarithmic density and temperature, and vertical velocity distribution at the 
vertical (xz) midplane are shown in panels (d)-(f) at $t=71.4\,\mathrm{min}$.}
\end{figure}
%========================================================

\begin{figure}[t]
\centering
\includegraphics[width=0.9\linewidth]{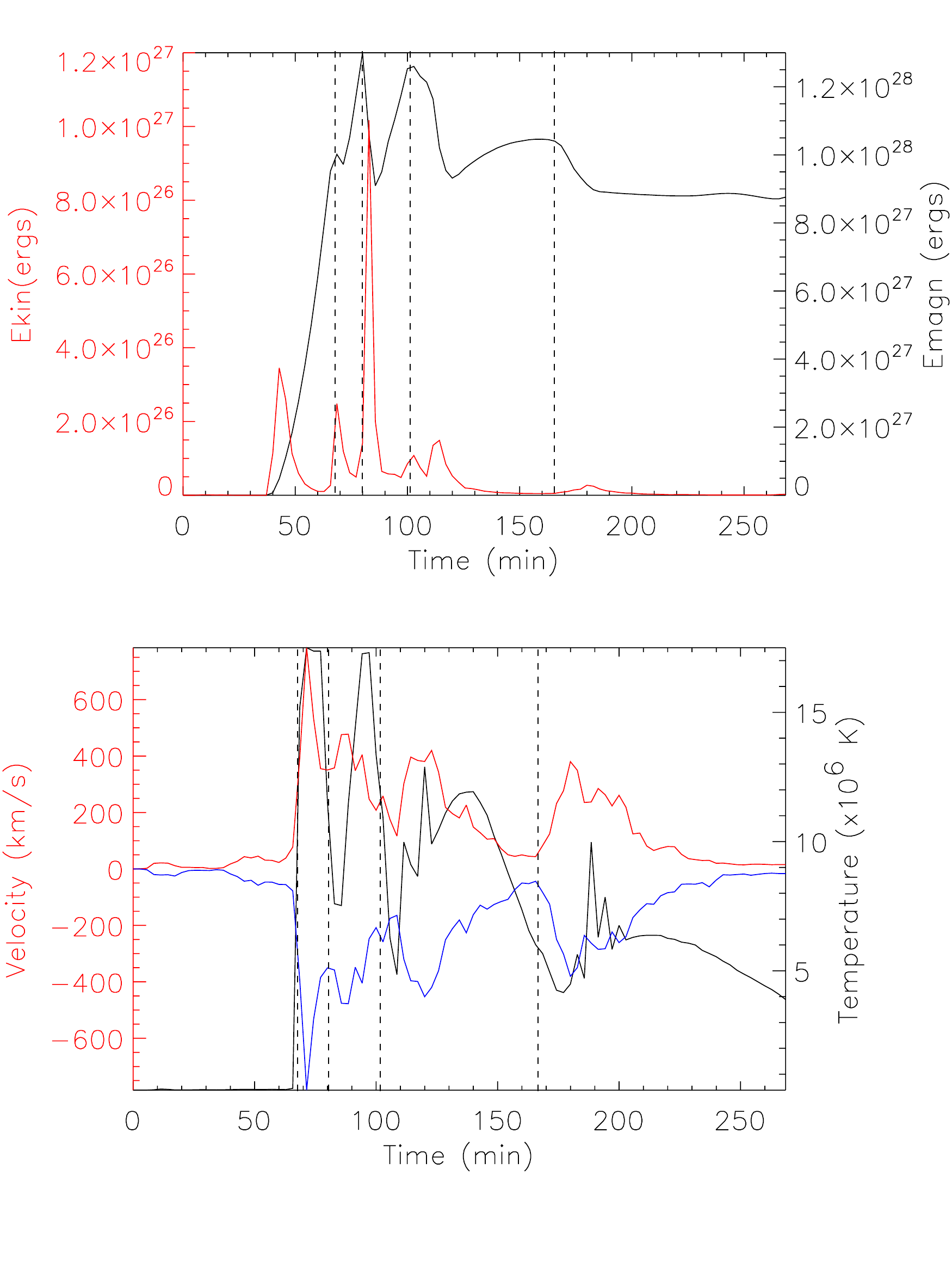}\caption{{\bf Top:} Temporal evolution of the magnetic (black line ) and
kinetic (red) energy above the photosphere. {\bf Bottom:} Temporal evolution of the maximum positive/negative (red/blue resepectively) $V_\mathrm{z}$ and
maximum temperature (black) above the photosphere.}
\end{figure}
%=========================================================
\begin{figure*}[t]
\centering
\includegraphics[width=0.9\linewidth]{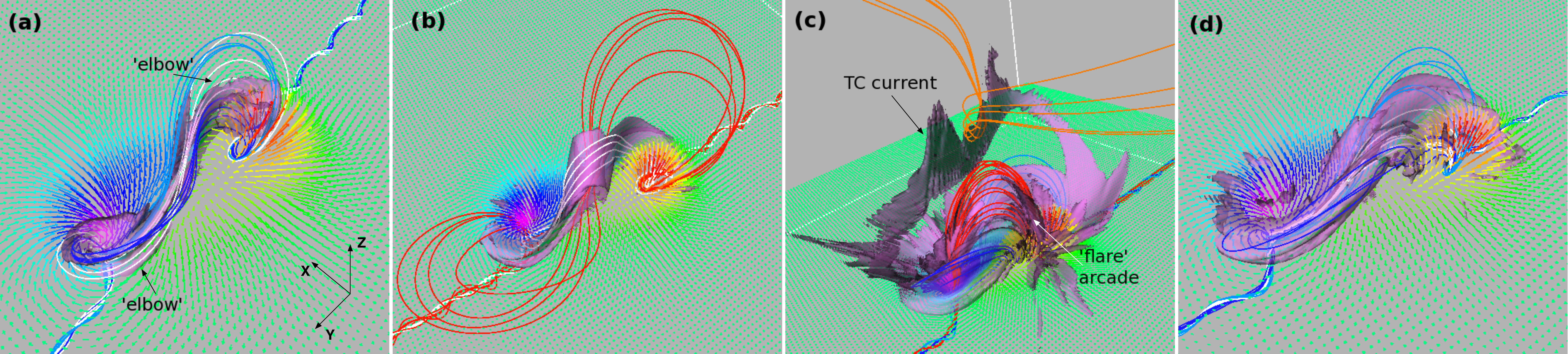}\caption{The sigmoid-to-arcade transformation during the first CME-like eruption
(panels a-c) and the reappearance of a sigmoid before the second eruption (panel d).
The value of the isosurface represents $\ge 50\%$ (panels a,b,d) and $\ge 25\%$
(panel c) of the maximum value of ${J\cdot B}/{\bf B}$ above the transition region. The horizontal slice shows the Bz distribution at the photosphere
(red/yellow;positive, blue/pink;negative within the range $[-450,450]\,\mathrm{G}$.)
Times are $t=65.7\,\mathrm{min}$, $t=68.5\,\mathrm{min}$, $t=71.4\,\mathrm{min}$ and $t=77.2\,\mathrm{min}$ for panels (a)-(d) respectively.}
\end{figure*}

Our experiment shows the recurrent eruptive behaviour of an EFR during 4.5 hours of its evolution.
Figure 1 shows the temperature distribution together with magnetic fieldlines, which have been traced from the 
footpoints of the sub-photospheric flux tube. Panel 1(a) shows the cool adiabatic expansion 
of the emerging field. The outermost fieldlines have already emerged into the corona forming an envelope (ambient) field for the 
magnetic flux, which continues to emerge from the solar interior. Part of this magnetic flux forms a new magnetic flux rope that erupts into the 
outer solar atmosphere in an ejective manner opening the envelope field (Panels 1(b),1(c)). There are at least four ejective 
eruptions, during the evolution (three of them are shown in Figure 1) of the EFR, displaying several common characteristics such as: a) the marked expansion 
of the erupting magnetized volume, b) the transport of dense (low-atmospheric) material to the high atmosphere, c) the explosive heating of the plasma underneath 
the erupting core of the field and d) the formation of a hot arcade with a cusp-like shape in the low atmosphere.

The formation of a new flux rope and the driving mechanism of its eruption in similar flux emergence experiments have been studied in 
previous simulations \citep[e.g.][]{maglon01, man04, arc08b, arc12}. Figure 2 (panels a-c) is a synoptic illustration of the formation 
and rise of the first new flux rope in the present experiment. At $t=62.8\,\mathrm{min}$ (2(a)), the emerging fieldlines (blue and cyan), which have 
previously undergone shearing, 
reconnect with each other to form two new sets of fieldlines: the white fieldlines that (will) constitute the central part of the developing flux rope and 
the yellow fieldlines that stay behind and do not erupt. The red fieldlines represent 
the envelope field. At $t=68.5\,\mathrm{min}$ (2(b)), the white fieldlines are traced 
from the center of the cross section of the erupting flux rope, at the vertical (xz) midplane. The blue and cyan fieldlines continue to reconnect, 
forming new fieldlines: the orange fieldlines, which are wrapping around the footpoints of the white fieldlines developing a
magnetic flux rope structure and the yellow fieldlines, which form an arcade. Notice that now the envelope fieldlines are stretched out 
so that they are about to reconnect in a tether-cutting manner underneath the flux rope. At $t=71.4\,\mathrm{min}$ (2(c)), the well-developed flux rope 
consists of straight (white) fieldlines along its axis and twisted fiedlines (blue/cyan/orange/red) around its axis.
An important result is that, at this stage of the evolution, the enevelope fieldlines reconnect in a tether-cutting manner at a strong and thin current layer 
under the rope.
This process triggers the ultimate ejective eruption of the flux rope towards the outer solar atmosphere. Therefore, in this case, the release of 
the downward tension of the
envelope fieldlines, which under certain conditions can halt the ejective eruption \citep[e.g.][]{arc12}, is not released by external reconnection 
with a pre-existing
coronal magnetic field \citep[e.g.][]{arc08b} in a break-out manner but by tether-cutting internal
reconnection underneath the erupting plasma.

Panel 2(d) shows that the eruption of the flux rope brings dense plasma from the low atmosphere into the corona. The heavy material ($\approx 1-2$ orders heavier 
than the background plasma) is accumulated 
at the dips of the twisted fieldlines of the flux rope. Panel 2(e) shows that the ejective phase of the eruption is characterized by the cool adiabatic 
expansion of the field and the heating of the plasma due to tether-cutting reconnection of the fieldlines. 
Panel 2(f) shows that high-speed ($\approx 800\,\mathrm{km\,s^{-1}}$) bi-directional jets are emitted vertically (upward/downwad) from the        
reconnection site underneath the rising flux rope.
The upward reconnection jet adds momentum to the rising flux rope and heats the
plasma inside the erupting volume. Therefore (panel 2e), cool (purple) and hot (yellow/red) plasma is found to be expelled together with the erupting
core of the flux rope. In the first two eruptions (which are faster, see also Figure 3), the top edge of the upward jet nearly reaches
to the core of the erupting field. Due to its high speed, it deforms the concave part of the twisted fieldlines that                
surrounds the erupting core from below. The downward reconnection jet collides with the top (flux pile-up) regime of the arcade 
forming a termination shock. As a result, the plasma is compressed locally at high temperatures, causing flaring of the arcade.  

Eventually, a second flux rope is formed at the upper photosphere/chromosphere, as new fieldlines emerge and undergo shearing and reconnection 
(such as the blue and cyan fieldlines in panels 2(a)-2(c)).
The innermost fieldlines of the arcade udergo the same process and, thus, they 
are incorporated in the formation of the flux rope. The outermost fieldlines of the arcade constitute a new envelope field for the second 
flux rope to rise into. Initially, the flux rope 
rises slowly (with a speed of about $40\,\mathrm{km\,s^{-1}}$ until $t=80\,\mathrm{min}$) and then it erupts in a fast-rise phase, accelerating to a velocity of $\approx 280\,\mathrm{km\,s^{-1}}$ 
in the next $90\,\mathrm{sec}$. 
The onset of acceleration is accompanied by tether-cutting reconnection of the new envelope's fieldlines.

The following eruptions occur in a slightly different manner. 
The magnetic field at the {\it center} of the EFR 
does not become strong enough to emerge above the photosphere with the standard $\Omega$-loop like configuration. 
On the contrary, it spreads out horizontally and the magnetic pressure increases in the 
vicinity of the two main polarities of the EFR. There, the magnetic fieldlines can emerge forming two magnetic lobes, similar to the emergence of 
a weakly twisted emerging field \citep[]{arc13}. The lateral expansion of the magnetic lobes bring their fieldlines into contact at the center of the EFR and 
reconnect, forming another flux rope. The eruption of the flux rope is driven by the Lorentz force, which is eventually enhanced by tether-cutting 
reconnection of the envelope fieldlines.

In summary, the tether-cutting reconnection following the first eruption restructures the
magnetic configuration in the solar atmosphere, so that the post-eruption state resembles the
pre-eruption state of the first eruptive event (Figure 2(a)). Eventually, emergence of new magnetic
flux and associated shearing, inject free magnetic energy into the system initiating a second
eruption very similar (homologous) to the first one.
Overall, the sequence of the homologous eruptions is attributed to a) ongoing flux emergence,
shearing and reconnection in the low atmosphere (e.g. up to the transition region) and b) to
the tether-cutting reconnection in the corona and the dynamical reconfiguration of the system
to a state similar to the initial one.

There are two differences in the topology of the field and the dynamics prior to the eruptions. Firstly, the
envelope field in the initial eruption consists of the uppermost emerging fieldlines, while in
the following eruptions it is formed by the fieldlines of the flare arcade. Secondly, the
formation of the erupting flux rope in the first two eruptions occurs due to shearing and reconnection
of the emerging field at the center of the AR. In the other two eruptions, the emerging field
develops two side magnetic lobes (see the work by \citet[]{arc13}), which eventually reconnect, forming
the magnetic flux rope.
Despite these differences, the key processes which are responsible for the driving
of the homologous eruptions in our experiment are similar in all events.

Figure 3 (top) shows the temporal evolution of the total magnetic and kinetic energy above the photosphere. The four eruptions start to occur at the 
times marked by the vertical dashed lines. In each eruption, magnetic energy drops and kinetic energy increases rapidly. The first increase of the 
kinetic energy ($t\approx 40\,\mathrm{min}$) corresponds to the initial emergence of the magnetic field. The second eruption starts while the lower part (i.e. where 
tether cutting reconnection of the envelope fieldlines occurs) of the first eruption is still within the numerical domain. 
For example, at $t\approx 83\,\mathrm{min}$ 
the tether cutting reconnection upflow (jet) of the first eruption originates at $z\approx 44\,\mathrm{Mm}$ and, thus, the 
corresponding total kinetic energy resembles the 
accumulated energy of the first two eruptions. After the first eruption, the decrease (increase) of the magnetic (kinetic) energy becomes lower with 
each of the following eruptions, which indicates that the CME-like eruptions become progressively less energetic. This is because there is a certain 
amount of flux and energy available to the system for driving the eruptions. This energy comes from the sub-photospheric magnetic flux tube. 
Due to dynamical emergence, the flux is eventually exchausted and, thus, the eruptions produced are less energetic.

Notice that the total magnetic energy increases after each eruption. This increase is important for the build-up of the neccessary amount of energy that is 
required for the onset of the successive eruptions. In the first and second eruption, the increase is due to the dynamical emergence and 
the shearing of the magnetic field around the center of the EFR. In the following eruptions, shearing is less pronounced but emergence continues to occur, which 
leads to the formation of the two magnetic lobes. Eventually, the system reaches a quasi-static equilibrium and the magnetic energy saturates.

Figure 3 (bottom) shows the temporal evolution of the vertical component of the velocity field (maximum and minimum value) and the temperature (maximum) above 
photosphere. Strong bi-directional flows are emitted during each eruption. These flows are the reconnection jets, which originate in the 
current sheet underneath the erupting flux rope. There is a very good correlation between the upflows and the downflows during the evolution of the 
system: for every upflow there is a corresponding downflow with the same almost magnitude. The fastest upflow occurs in the first eruption ($\approx 
800\,\mathrm{km\,s^{-1}}$). In the other eruptions, the associated jets run with progressively lower speeds, within the range $450-350\,\mathrm{km\,s^{-1}}$. 
The measured velocity 
at the center of the erupting flux ropes, in the last snapshot before they exit the domain and while they are accelerated, is 
between $\approx 200-250\,\mathrm{km\,s^{-1}}$. This indicates that indeed the reconnection jets can add momentum to the erupting plasma volume. 

The evolution of the temperature shows that each eruptive event is followed by intense plasma heating. In a similar manner to the temporal evolution 
of $V_\mathrm{z}$, the maximum heating decreases over time. It drops from $17\,\mathrm{MK}$ after the first eruption to about $10\,\mathrm{MK}$ in the 
last eruptive event. Therefore, 
the heating decreases when the eruption is less energetic. The intense heating occurs at the apex of the cusp-like arcade underneath the 
reconnection site of the produced jets.
Therefore, each eruption is followed by a flaring episode. The time offset between the onset of the eruption 
and the heating of the arcade is determined by the 
time that the downward reconnection jet takes to reach the arcade and heat it.

The recurrent CME-like eruptions are associated with the appearance of sigmoidal structures in the EFR. Figure 4 shows the 
three-dimensional visualization of the isosurface ${J\cdot B}/{\bf B}$, where J is the electric current and B is the magnetic field strength. 
At $t=65.7\,\mathrm{min}$ (panel (a)), the isosurface and the fieldlines that are traced from this structure adopt an overal S-like shape associated with the twist and writhe 
of the rising magnetic field. Some of these fieldlines (white) are dipped at the center of the sigmoid and expand towards its two ends (``elbows''). 
There are also sheared fieldlines (blue/cyan) that surround only one of the ``elbows'' and, thus, they make only a half turn along the sigmoid. 
As these fieldlines come closer together, the current becomes large at their interface (along the PIL) developing the sigmoidal structure. 
Eventually, the blue and cyan fieldlines will reconnect around the center of the EFR forming a new flux rope. Due to reconnection, we expect that the  
sigmoid's fieldlines will be heated to high temperatures. At this stage of the simulation, we find that the dense plasma of the sigmoidal structure has an 
average temperature of the $O(5\,\times 10^{5}) {K}$ over the integrated volume (i.e. above the transition region) of the isosurface and, thus, it may 
appear as a bright S-shaped structure at the solar atmosphere.

At $t=68.5$ (panel (b)), the sigmoid develops a loop-like structure at its center. This is the current-currying flux rope that starts to rise, leading to the 
partial eruption of the sigmoid. The (white) fieldlines, which have been traced from the loop, belong to the erupting flux rope. The red fieldlines have 
been traced from the center of the sigmoidal isosurface underneath the flux rope.
The flaring of the plasma, which follows the eruption, starts from this central area of the sigmoid. 
As the eruption proceeds (panel (c)), a new thin and curved current layer is formed underneath the erupting plasma. This 
is where tether-cutting reconnection occurs (TC current). 
Thus, fieldlines (orange), which were previously part of the enevelope field, now reconnect adopting a V-shape at 
the center of the current layer. 
Under the TC current, the downward reconnected fieldlines (red) 
form the ``flare'' arcade, which is heated to more than $15\,\mathrm{MK}$. Thus, the bright arcade does not appear to be part of the 
pre-flare sigmoid but it is a new structure associated with the tether-cutting reconnection. The (blue/cyan) fieldlines are traced from the lower-lying 
remnants of the sigmoid, which temporarily fades away. However, shortly after the first eruption (panel (d)), the dynamical emergence 
and the shearing of the field continue to operate, leading to reappearance of a sigmoidal structure (blue/cyan/white fieldlines have been traced 
in a similar manner to panel a).
A similar process (i.e. sigmoid-to-arcade conversion) is repeated during the following eruptions. 

We have performed a preliminary parametric study, which shows that the parameters of the system may affect its evolution and dynamics. 
For instanse, for larger values of plasma $\beta$ or $\lambda$ of the sub-photospheric flux tube the eruptions occur at later times. In adittion, we find that 
for an initial field strength of the tube that correrponds to $\beta \geq 39$, the eruptions are confined. 
The ejection of helicity and the reformation of sigmoids for a wider range of parameters 
will be presented in forthcoming work.

\acknowledgments
The simulations were performed on the STFC and SRIF funded
UKMHD cluster, at the University of St Andrews. The authors acknowledge support by EU (IEF-272549 grant) and 
the Royal Society. 
V.A and A.W.H are grateful
for in-depth discussions during the ISSI workshop: ``Magnetic flux emergence in the solar atmosphere'' in Bern.

\bibliographystyle{apj}

\clearpage

\end{document}